\documentclass[10pt]{article}

\usepackage{amsfonts}
\usepackage{amsmath}
\usepackage{amssymb}
\usepackage{enumerate}

\usepackage{graphicx}
\usepackage{sidecap}

\usepackage{bm}
\usepackage{color} 
\usepackage{multicol}
\usepackage{multirow}

\usepackage{amsthm}
\theoremstyle{definition}
\newtheorem{theorem}{Theorem}

\newtheorem{example}[theorem]{Example}

\newcommand{\R}{\mathbb{R}}

\usepackage[boxed,vlined,ruled,linesnumbered]{algorithm2e}

\usepackage{fancyhdr}

\usepackage[
top    = 3.00cm,
bottom = 3.00cm,
left   = 2.50cm,
right  = 2.50cm]{geometry}

\title{\textbf{An implementation of CAD in Maple} \\ \textbf{utilising problem formulation, equational} \\ \textbf{ constraints and truth-table invariance}}
\author{Matthew England}
\date{Department of Computer Science, University of Bath, Bath, UK \\ \texttt{M.England@bath.ac.uk} }

\begin{document}

\maketitle

\pagestyle{fancy}
\lhead{M.~England}
\rhead{An implementation of CAD in Maple utilising McCallum projection}

\begin{abstract}
Cylindrical algebraic decomposition (CAD) is an important tool for the investigation of semi-algebraic sets, with applications within algebraic geometry and beyond.  We recently reported on a new implementation of CAD in \textsc{Maple} which implemented the original algorithm of Collins and the subsequent improvement to projection by McCallum.  Our implementation was in contrast to \textsc{Maple}'s in-built CAD command, based on a quite separate theory.  Although initially developed as an investigative tool to compare the algorithms, we found and reported that our code offered functionality not currently available in any other existing implementations.
One particularly important piece of functionality is the ability to produce order-invariant CADs.  This has allowed us to extend the implementation to produce CADs invariant with respect to either equational constraints (ECCADs) or the truth-tables of sequences of formulae (TTICADs).  This new functionality is contained in the second release of our code, along with commands to consider problem formulation which can be a major factor in the tractability of a CAD.  

In the report we describe the new functionality and some theoretical discoveries it prompted.  We describe how the CADs produced using equational constraints are able to take advantage of not just improved projection but also improvements in the lifting phase.  We also present an extension to the original TTICAD algorithm which increases both the applicability of TTICAD and its relative benefit over other algorithms.  The code and an introductory \textsc{Maple} worksheet / pdf demonstrating the full functionality of the package are freely available online.
\end{abstract}

\noindent This work is supported by EPSRC grant EP/J003247/1.

\section{Introduction} \label{SEC_Intro}

This report is the second, following \cite{England13}, to describe our implementation of cylindrical algebraic decomposition (CAD) in \textsc{Maple}. This report describes the functionality added to the second release of the code.  This includes the ability to: produce CADs invariant with respect to equational constraints following \cite{McCallum99}, produce truth-table invariant CADs (TTICADs) following \cite{BDEMW13}, and derive different formulations of the input with heuristics to pick the best following \cite{BDEW13}.  The introduction continues with a brief background on CAD and a summary of the findings of \cite{England13}.  Then in Section \ref{SEC_PCAD} we describe the new functionality in more detail.

As with \cite{England13}, the implementation prompted some theoretical discoveries which are also described in this report.  In Section \ref{SEC_EC} we explain how the use of equational constraints can not only lead to the improved projection described in \cite{McCallum99}, but also improved lifting.  Then in Section \ref{SEC_TTICAD} we present an extension to the original TTICAD algorithm of \cite{BDEMW13}, allowing it to be applied to a wider variety of input formulae, including formulae which could not be tackled by equational constraints alone.  We demonstrate that this increases not just the applicability of TTICAD, but its relative benefit over other algorithms.

The second release of the \texttt{ProjectionCAD} code and an introductory \textsc{Maple} worksheet / pdf demonstrating the full functionality of the package are freely available online at \texttt{http://opus.bath.ac.uk/35636/}. 

\newpage

\subsection{Background on CAD}

A cylindrical algebraic decomposition (CAD) is a decomposition of $\R^n$ into cells, constructed with respect to an input, usually either polynomials or formulae, in $n$ variables. Each cell describes a semi-algebraic set and the cells are cylindrically arranged, meaning the projection of any two cells is either equal or disjoint.  A CAD is sign-invariant if the input polynomials have constant sign on each cell.  Such a CAD allows for the solution of many problems defined by the polynomials.  Collins provided the definition and first algorithm \cite{Collins75, ACM84I}, motivated as a tool for quantifier elimination in real closed fields.
Since their discovery they have found many other applications ranging from robot-motion planning \cite{SS83II, Davenport86} to simplification technology \cite[etc.]{BD02, DBEW12}

Collins' algorithm has two phases.  The first, \textit{projection}, applies a projection operator repeatedly to a set of polynomials, each time producing another set 
in one fewer variables.  Together these sets contain the {\em projection polynomials}.  The second phase, \textit{lifting}, then builds the CAD incrementally from these polynomials.  First $\R$ is decomposed into cells which are points and intervals corresponding to the real roots of the univariate polynomials.  Then $\R^2$ is decomposed by repeating the process over each cell using the bivariate polynomials at a sample point of the cell.  The output for each cell consists of {\em sections} of polynomials (where a polynomial vanishes) and {\em sectors} (the regions between these). Together these form the  {\em stack} over the cell, and taking the union of these stacks gives the CAD of $\R^2$.  This process is repeated until a CAD of $\R^n$ is produced.  The projection operator must be chosen in order to conclude that the CAD of $\R^n$ produced using sample points in this way is sign-invariant.
The output of a CAD algorithm depends on the ordering of the variables.
In this paper we usually work with polynomials in $\mathbb{Z}[x_1,\ldots,x_n]$ with the variables, ${\bf x}$, listed in ascending order; (so we first project with respect to $x_n$ and so on until we have univariate polynomials in $x_1$).  
The \textit{main variable} of a polynomial, ${\rm mvar}(f)$, is the greatest variable present with respect to the ordering.  


Since Collins published the original algorithm there has been much research into improvements with a summary of the developments over the first twenty years given by \cite{Collins98}.  A key area of work was in the definition of projection operators to use in the first phase.  McCallum defined an operator \cite{McCallum88, McCallum98} which usually produces far fewer polynomials than Collins' and thus an algorithm which uses it produces a CAD with less cells in less time.  However, unlike Collins' operator, McCallum's cannot be applied universally.  McCallum defined the notion of input which is  well-oriented \cite{McCallum98} to describe when his operator could be used.

\textsc{Qepcad} \cite{Brown03b} is a dedicated CAD implementation: a command-line program 
which implements both McCallum's projection operator and Hong's modification of Collins' operator \cite{CH91}).  
In 2009 a radically different approach was presented \cite{CMXY09} where instead of projection and lifting, a cylindrical decomposition of complex space is first produced which is then refined to a CAD of real space.  That algorithm is distributed with {\sc Maple} where it is part of the \texttt{RegularChains} library \cite[etc.]{MorenoMaza99}.

\subsection{The \texttt{ProjectionCAD} package}

\texttt{ProjectionCAD} is a \textsc{Maple} package developed at The University of Bath to implement CAD via projection and lifting.  It was initially developed in order to study the differences between the traditional approach to CAD and the new approach in \cite{CMXY09} using regular chains.  In \cite{England13} we described our implementation of both McCallum's and Collin's algorithms to produce sign-invariant CADs, and how we made use of commands from the \texttt{RegularChains} library to give output in the same format as \cite{CMXY09}.

\texttt{ProjectionCAD} is a theoretical tool, unoptimised, and so does not compete particularly well on timings for sign-invariant CADs (see \cite[Section 4]{England13} for a comparison).  However, it is currently the only implementation which can produce order-invariant CADs, (CADs with the stronger property that each polynomial has constant order of vanishing on each cell).  This feature is essential for the extensions to the package described in Section \ref{SEC_PCAD}, which in turn allow this implementation to compete with the others.

\texttt{ProjectionCAD} is also the only implementation of delineating polynomials \cite{McCallum98, Brown05} which modify Collins' original lifting algorithm to allow McCallum's projection operator to be applied more widely.  Without these, the condition of well-orientedness has to be strengthened to thus restricting when the improved projection operator may be used.  \textsc{Qepcad} does not make use of delineating polynomials and so there are examples where \texttt{ProjectionCAD} can produce CADs which \textsc{Qepcad} cannot (see \cite[Section 2]{England13}.  However, \textsc{Qepcad} has implemented many of the the ideas in \cite{Brown05} in order to minimise the effect of this.

\section{New functionality in \texttt{ProjectionCAD} V2} 
\label{SEC_PCAD}

In this section we describe the new functionality available in the second release of the \texttt{ProjectionCAD} package.  This adds to the work in the first release (described in \cite{England13}) for producing sign and order-invariant CADs.

\subsection{Equational constraints}
\label{SUBSEC_EC}

An \textbf{equational constraint} is an equation logically implied by a formula.  Their use in CAD is based on the observation that the formula will be false for any cell in the CAD where the equation is not satisfied.  Hence, instead of a CAD sign-invariant with respect to all the polynomials, we could use one that is: (a) sign-invariant with respect to the equational constraint; and (b) sign-invariant with respect to the other constraints only in those cells over the sections of the equational constraint.  Such a CAD is said to be {\bf invariant with respect to an equational constraint}.  This observation was made in \cite{Collins98} with the first detailed approach given by \cite{McCallum99} where a new projection operator was presented to implement the idea.

\subsubsection*{The background theory}

Let $P$ be the McCallum projection operator.  Informally this is applied to a set of polynomials to produce the coefficients, discriminant and cross resultants (full details of the theory are presented in \cite{McCallum88, McCallum98} and the implementation in \texttt{ProjectionCAD} is discussed in \cite{England13}).  We note that most implementations including \texttt{ProjectionCAD} make trivial simplifications such as removal of constants, exclusion of polynomials that are identical to a previous entry (up to constant multiple), and only including those coefficients which are necessary for the theory to hold.  
This operator may be used for CAD provided the input polynomials are \textbf{well oriented}.  This is defined to mean that every projection polynomial has a finite number of nullification points, i.e. if the main variable is $x_k$ then $f(\alpha,x_k)=0$ for at most a finite number of $\alpha \in \R^{k-1}$.

Given a problem with an equational constraint $f$ McCallum suggested a reduced projection operator $P_f$.  Informally, this consists of the coefficients and discriminant of $f$ together with the resultant of $f$ taken with each of the other polynomials.  
Formally, we must take more care leading to an operator $P_F$ where $F$ is a squarefree basis for the primitive part of $f$ (see \cite{McCallum99} for the full details of the theory).
This improved projection operator is used for the first projection, reverting to $P$ for subsequent projections.  To use this new operator in place of the original McCallum operator the polynomials in the CAD algorithm the input must again satisfy the well-orientedness condition

\subsubsection*{The implementation in \texttt{ProjectionCAD}}

Algorithm \ref{alg_ECCAD} describes our implementation in \texttt{ProjectionCAD} which builds a CAD of $\R^n$ with respect to an equational constraint.  While the projection phases follows the work of \cite{McCallum99} exactly the lifting phase is somewhat different, as discussed in detail in Section \ref{SEC_EC}.  The algorithm uses the sub-algorithms: \texttt{CADFull} (Algorithm 3 in \cite{England13}) to build an order-invariant CAD for the projection polynomials not in the main variable; \texttt{LiftingSet} (Algorithm \ref{alg_LSet}) to calculate the set of polynomials to use for the final lift; and \texttt{CADGenerateStack} (Algorithm 4 in \cite{England13}) to then lift over cells with respect to these polynomials.   

\begin{algorithm} \caption{ECCAD} \label{alg_ECCAD}
\DontPrintSemicolon
\SetKwInOut{Input}{Input}\SetKwInOut{Output}{Output}
\Input{
$\bullet$ A polynomial $f \in \R[x_n,\ldots,x_1]$. 
$\bullet$ A set of polynomials $G \subset \R[x_n,\ldots,x_1]$. 
}
\Output{Either: 
$\bullet$ a CAD of $\R^n$, sign-invariant with respect to $f$ and, when $f=0$, also with respect to $G$, or; $\bullet$ {\bf FAIL} if the input is not well-oriented.}
\BlankLine

Set $E := \{f\}$.\;
Compute $F$, the finest squarefree basis for the primitive parts of E.\;
\If{$n=1$}{
\Return The CAD of $\R$ formed by the decomposition of the real line according to the real roots of the polynomials in $F$.\;
}
Set $A := G \cup E$.\; 
Compute the set $C$ of contents of the elements of $A$.\; 
Compute the set $B$, the finest squarefree basis for the primitive parts of $A$.\;
Construct the projection set $\mathfrak{P} := C \cup P_{F}(B)$. \label{step_EC_mfP} \;
Attempt to construct a lower-dimensional CAD by calculating $D := {\tt CADFull}(\mathfrak{P}, \rm{proj=McCallum}, fcad=\rm{true})$. \label{step_EC_CF} \;
\If{ \rm{\texttt{CADFull} warns about potential failure} }{
\Return {\bf FAIL}.\;
}
\For{each cell $c \in \mathcal{D}$}{
Set $L_c := \texttt{LiftingSet}(c,A,E)$. \;
\If{$L_c =$ {\bf FAIL}}{
\Return {\bf FAIL}.\; }
Set $S_c := \texttt{CADGenerateStack}(c,L_c)$.\;
}
\Return $\bigcup_c S_c$.\;
\end{algorithm}

\begin{algorithm} \caption{LiftingSet} \label{alg_LSet}
\DontPrintSemicolon
\SetKwInOut{Input}{Input}\SetKwInOut{Output}{Output}
\Input{
$\bullet$ A cell $c$ from a CAD of $\R^{n-1}$. 
$\bullet$ A set of polynomials $A \subset \R[x_n,\ldots,x_1]$. \\
$\bullet$ A subset of polynomials $E \subset A$.  \\
}
\Output{Either: $\bullet$ a set of polynomials $L \subset \R[x_n,\ldots,x_1]$ to use for lifting over the cell, or;\\ $\bullet$ {\bf FAIL} if the input is not well-oriented.}
\BlankLine
Set $L := \{\}$.\;
\eIf{\rm{any polynomial in} $E$ \rm{is nullified on} $c$ \label{step_EC_WO} }{
\eIf{$\dim(c)>0$}{
calculate ExclP$_E(A):=P(A \setminus E) \setminus P_E(A)$. \label{step_EC_excl1} \;
\eIf{\rm{ExclP}$_E(A)$ \rm{is empty or contains only constants} }{
$L := L \cup A_i$. \label{step_EC_addB1}  \;}{
\Return {\bf FAIL}.\;}
}{
$L := L \cup A_i$. \label{step_EC_addB2} \;}
}{
$L := L \cup E_i$.\;
}
\end{algorithm}

It is well documented that CADs invariant with respect to equational constraints usually have fewer cells than those which are sign-invariant for all polynomials.  Example \ref{ex_EC1} gives a simple demonstration of this.

\begin{example}
\label{ex_EC1}
Assume the variable ordering $y>x$.  Consider the polynomials 
\[
f = x^2+y^2-4, \qquad \mbox{and} \qquad g= xy-1
\] 
which define the circle and hyperbola in Figure \ref{fig_exEC}, and the problem $f=0 \wedge g<0$.  We assume the variable ordering $y>x$.  The problem could be solved by means of a full sign-invariant CAD for $\{f,g\}$ which \texttt{ProjectionCAD} would produce using 83 cells.  However, it is more efficient to produce a CAD invariant with respect to the equational constraint $f$, which using Algorithm \ref{alg_ECCAD} in \texttt{ProjectionCAD}, has 53 cells.  

The induced CADs of the real line have 15 and 13 cells respectively.  The difference is that the full CAD identifies the origin on the real line, corresponding to the asymptote of the hyperbola and arising from the inclusion of the coefficients of $g$ is the projection polynomials.  The CAD using the equational constraint only considers the behaviour of $g$ on cells where $f=0$, (by including the resultant of $f$ and $g$ in the first projection but not the coefficients of $g$ individually) and thus the origin is not identified. 

\begin{SCfigure}
\caption{Plots of the curves used in Example \ref{ex_EC1}.}
\label{fig_exEC}
\centering
\includegraphics[height=5cm]{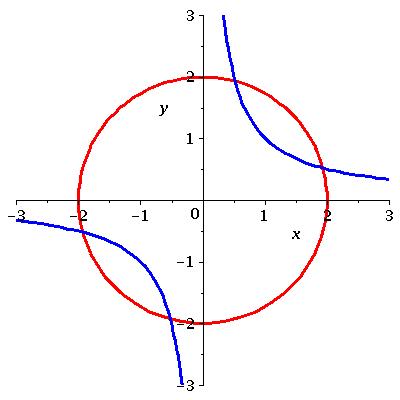}
\end{SCfigure}

\end{example}

As we discuss further in Section \ref{SEC_EC}, Algorithm \ref{alg_ECCAD} actually offers some subtle improvements to the original algorithm described in \cite{McCallum99} and implemented in \textsc{Qepcad}.  For the example above this manifests in the cell count of 53 produced by \texttt{ProjectionCAD} comparing to a cell count of 69 with \textsc{Qepcad} as described in Example \ref{ex_EC-small}.

\subsection{TTICAD}

Given a sequence of formulae, a \textbf{truth table invariant CAD} (TTICAD) is a CAD such that each formula has constant truth value (true or false) on each cell.  The idea of using truth invariance was introduced in \cite{Brown98} for use in simplifying sign-invariant CADs.  Of course, a sign-invariant CAD is itself truth-invariant but by focusing on the second property certain cells can be merged to give a smaller CAD.  Recently, in \cite{BDEMW13} an algorithm was presented allowing the efficient construction of TTICADs (without having to first build a sign-invariant CAD).  

\subsubsection*{The background theory}

The algorithm presented in \cite{BDEMW13} makes use of the theory of equational constraints.  It assumes that each formula in the sequence has a designated equational constraint (for that formula only).  Then a new projection operator was defined which, informally, included the following polynomials.
\begin{itemize}
\item For each formula, those obtained by applying McCallum's projection operator for equational constraints (see \cite{McCallum99} and Section \ref{SUBSEC_EC} above).
\item For each formula, the polynomials obtained by taking the resultant of the designated equational constraint and each of the other polynomials.
\item The cross-resultants of the set of designated equational constraints from each formula.
\end{itemize}
This new projection operator is used for the first projection, with the original McCallum projection operator used for subsequent projections.  Constructing the projection set this way ensures that the CAD produced is sign-invariant with respect to equational constraints, and that for cells where an equational constraint is zero, the other polynomials from that formula are sign-invariant.

As with the other projection operators, this one is only valid for use when the input sets satisfy a condition.  The well-orientedness condition described above applied to the polynomials in the input would be sufficient, but instead, a finer condition is used, discussed further below.

\subsubsection*{The implementation in \texttt{ProjectionCAD}}

In \cite{BDEMW13} an algorithm was provided and verified for the approach above.  In Algorithm \ref{alg_TTICAD} we present the algorithm used in \texttt{ProjectionCAD}.  This extends the algorithm in \cite{BDEMW13} so that it may be applied to sequences of formulae in which not every formula has a designated equational constraint.   

First we formally define the projection operator used.  Let $\mathcal{A} = \{ A_i\}_{i=1}^t$ be a list of irreducible bases $A_i$
and let $\mathcal{E} = \{ E_i \}_{i=1}^t$ be a list of non-empty subsets $E_i \subseteq A_i$.  Put $A = \bigcup_{i=1}^t A_i, E = \bigcup_{i=1}^t E_i$.  (We use the convention of uppercase Roman letters for sets and calligraphic letters for sequences).

Then define the {\bf reduced projection of $\mathcal{A}$ with respect to $\mathcal{E}$}, denoted by $P_{\mathcal{E}}(\mathcal{A})$, as follows:
\[
P_{\mathcal{E}}(\mathcal{A}) := \textstyle{\bigcup_{i=1}^t} P_{E_i}(A_i) 
\cup {\rm RES}^{\times} (\mathcal{E}) 
\]
where 
\begin{align*}
P_{E}(A) &= P(E) \cup \left\{ {\rm res}_{x_n}(f,g) \mid f\in E, g \in A, g \notin E \right\}; 
\\
{\rm RES}^{\times} (\mathcal{E}) 
&= \{ {\rm res}_{x_n}(f,\hat{f}) \mid \exists \, i,j \,\, \mbox{such that } 
  f \in E_i, \hat{f} \in E_j, i<j, f \neq \hat{f}  \}.
\end{align*}
We say that $\mathcal{A}$ is {\bf well oriented with respect to $\mathcal{E}$} if whenever $n > 1$: every polynomial $f \in E$ is nullified by at most a finite number of points in $\R^{n-1}$, and $P_{\mathcal{E}}(\mathcal{A})$ is well-oriented in the original sense.
Algorithm \ref{alg_TTICAD} shows how the projection operator may be applied more widely. 

The definitions above are the same as \cite{BDEMW13} but their use in Algorithm \ref{alg_TTICAD} differs from their use in the algorithm presented in \cite{BDEMW13}.  This is discussed further in Section \ref{SEC_TTICAD}.  Experimental results in \cite{BDEMW13} show that TTICAD offers huge benefits compared to sign-invariant CADs constructed with the same technology (\texttt{ProjectionCAD}).  It also demonstrates that the TTICAD theory allows \texttt{ProjectionCAD} to compete with the existing state of the art CAD technology.

\begin{SCfigure}
\caption{Plots of the curves used in Example \ref{ex_TTICAD1}.  The solid circle is $f_1$, the solid hyperbola $g_1$, the dashed circle $f_2$ and the dashed hyperbola $g_2$.}
\label{fig_exTTI}
\centering
\includegraphics[height=4cm]{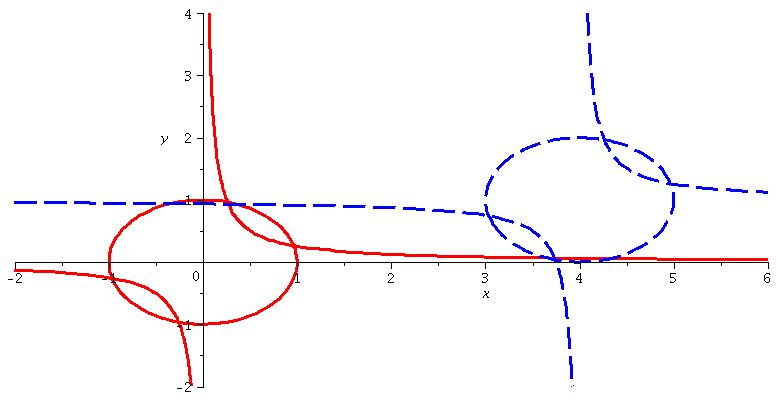}
\end{SCfigure}

\begin{example}
\label{ex_TTICAD1}
Assume the variable ordering $y>x$.  Consider the polynomials:
\begin{align*}
f_1 = x^2+y^2-1 \qquad \qquad \qquad & g_1 = xy - \tfrac{1}{4} \\
f_2 = (x-4)^2+(y-1)^2-1  \quad & g_2 = (x-4)(y-1) - \tfrac{1}{4}
\end{align*}
which are plotted in Figure \ref{fig_exTTI}, and the formula
\[
\Phi= \left(f_1 = 0 \land g_1 < 0 \right)\lor \left( f_2 = 0 \land g_2 < 0   \right).
\]
Using \textrm{\texttt{ProjectionCAD}} a full sign-invariant CAD with respect to the polynomials can be constructed using 317 cells while a TTICAD can be produced using only 105 cells. 
The problem could also be tackled using the theory of equational constraints alone, (by declaring the implicit equational constraint $f_1f_2=0$).  Using Algorithm \ref{alg_ECCAD} in \texttt{ProjectionCAD} produces a CAD using 145 cells.  There are more cells than the TTICAD because the projection set will include polynomials relating to the intersection of $f_1$ with $g_2$ and $f_1$ with $g_2$ which are ignored by the TTICAD. 
This example was worked through and discussed in detail in \cite{BDEMW13}. 
\end{example}

\begin{algorithm}[h!]
\label{alg_TTICAD}
\DontPrintSemicolon
\SetKwInOut{Input}{Input}\SetKwInOut{Output}{Output}
\Input{A list of quantifier-free formulae $\Phi = \{ \phi_i \}_{i=1}^t$ in variables $x_1,\ldots,x_n$. Each $\phi_i$ may or may not have a designated equational constraint $f_i = 0$.
}
\Output{Either 
$\bullet$ $\mathcal{D}:$ A TTICAD of $\R{}^n$ for $\Phi$; or {\ } 
$\bullet$~{\bf FAIL}: If $\Phi$ is not well oriented
}
\BlankLine
\For{$i = 1 \dots t$}{
Extract the set $A_i$ of polynomials in $\phi_i$. \; 
\eIf{$\phi_i$ has a designated equational constraint $f_i$ \label{step_T1} }{
set $E_i := \{f_i\}$. \label{step_T2}  \;
}{
set $E_i := A_i$. \label{step_T3}  \;
}
Compute the finest squarefree basis $F_i$ for the primitive parts of $E_i$.\;
} 
Set $F := \cup_{i=1}^t F_i$.\;
\If{$n=1$}{
\Return The CAD of $\R$ formed by the decomposition of the real line according to the real roots of the polynomials in $F$.\;
}
\For{$i = 1 \dots t$}{
Compute the set $C_i$ of contents of the elements of $A_i$.\; 
Compute the set $B_i$, the finest squarefree basis for the primitive parts of $A_i$.\;
} 
Set $C := \cup_{i=1}^t C_i$,
$\mathcal{B} := (B_i)_{i=1}^t$ and 
$\mathcal{F} := (F_i)_{i=1}^t$. \;
Construct the projection set: $\mathfrak{P} := C \cup P_{\mathcal{F}}(\mathcal{B})$. \;
Attempt to construct a lower-dimensional CAD by calculating $D := {\tt CADFull}(\mathfrak{P}, \rm{proj=McCallum}, fcad=\rm{true})$\;

\If{ \rm{\texttt{CADFull} warns about potential failure} }{
\Return {\bf FAIL}.\;
}
\For{each cell $c \in \mathcal{D}$}{
Set $L_c := \texttt{LiftingSet}(c,A_i,E_i)$. \;
\If{$L_c =$ {\bf FAIL}}{
\Return {\bf FAIL}.\; }
Set $S_c := \texttt{CADGenerateStack}(c,L_c)$.\;
}
\Return $\bigcup_c S_c$.\;
\caption{{TTICAD}}
\end{algorithm}

\subsection{Problem formulation and heuristics}

The algorithms described above and in \cite{England13} can have different results for the same problem depending on how the problem is formulated.  For example, which variable ordering is used, which equational constraint is designated, how a formula is broken down into sub-formulae for TTICAD.  

In \cite{BDEW13} these issues were studied in detail and heuristics were developed for choosing the formulations.  The heuristics were based on two measures of CAD complexity which can be applied to the the projection polynomials.  The first measure, introduced in \cite{DSS04}, is the sum of total degrees of the each monomial in each polynomial, abbreviated to \texttt{sotd}.  The second introduced in \cite{BDEW13} in the number of distinct real roots of the univariate projection polynomials (the number of sections in the induced CAD of $\R$), abbreviated to \texttt{ndrr}.  
Both measures have been implemented in \texttt{ProjectionCAD}, the second by first taking a square free basis of the polynomials and then using \textsc{Maple}'s implementation of Sturm chains.  Heuristic algorithms have also been introduced to consider possible formulations and pick the best based on the values of these measures.  The user can specify to use either measure, a combination for successively breaking ties or a weighted average.  There is also implementation of the greedy algorithm for choosing variable orderings given in \cite{DSS04}, and the ability to specify variable blocks and then let the algorithm pick the best ordering which respects these blocks (as required when using CAD for quantifier elimination).

\section{Improved lifting with equational constraints} 
\label{SEC_EC}

The implementation of CAD with respect to an equational constraint described in Section \ref{SUBSEC_EC} offers some subtle improvements to the original work of \cite{McCallum99} and implementations based on this such as \textsc{Qepcad}.  

As discussed, the approach rests on the use of an improved projection operator $P_f$ to build a CAD, taking the place of either the Collins or original McCallum projection operator which construct full sign-invariant CADs.  When defined in \cite{McCallum99} the author only discussed how this would improve the projection phase of CAD, by creating fewer projection polynomials.  The only modification to the lifting phase described was the need to check the well-orientedness condition, as is also the case if using the original McCallum operator.  

Algorithm \ref{alg_ECCAD} makes further changes to the lifting phase, which although subtle can have a significant effect on the CADs produced.  We note that these ideas are also used in Algorithm \ref{alg_TTICAD} for producing TTICADs.

\subsection{A finer check for well-orientedness}
\label{SUBSEC_finer}

Theorem 2.3 of \cite{McCallum99} is the key result that justifies the use of the improved projection operator for equational constraints.  The theorem allows us to confirm that the cells in the outputted CAD which are over the sections of the equational constraint are sign-invariant with respect to the other polynomials.  The theorem only holds when the equational constraint has a finite number of nullification points, a condition guaranteed by the input being well-oriented.  

However, the standard well-orientedness condition is actually stronger than necessary for this situation.  It requires that all projection polynomials have a finite number of nullification points, including the non-equational constraints.  Algorithm \ref{alg_ECCAD} does not check the condition for every projection polynomial.  When using \texttt{CADFull} in step \ref{step_EC_CF} the condition is checked for all projection polynomials not in the main variable, while in step \ref{step_EC_WO} of Algorithm \ref{alg_LSet} the condition is checked for the equational constraint.  However, the condition is never checked for the non-equational constraints (in the main variable) as this is not required for the theory of the projection operator to hold.  
\textsc{Qepcad} appears to check for nullification of all projection polynomials, returning errors even if the theory does allow for the outputted CAD to have the requested properties.  Although it has detailed checks to avoid unnecessary errors \cite{Brown05} Example \ref{ex_EC-finer} demonstrates that these false errors may still occur, arising from this situation.

\begin{example}
\label{ex_EC-finer}
Consider the polynomials 
\[
f = x+y+z+w, \qquad g = zy - x^2w
\]
and the formula $f=0 \wedge g<0$.  We could analyse the truth of the formula using a full sign-invariant CAD for $\{f,g\}$ which both \texttt{ProjectionCAD} and \textsc{Qepcad} would produce with 557 cells.  However, it is far more efficient to make use of the equational constraint.  Using Algorithm \ref{alg_ECCAD} \texttt{ProjectionCAD} produces a CAD with 165 cells.  Declaring the equational constraint in QEPCAD results in a CAD with 221 cells (the higher number due to the issues discussed in subsection \ref{SUBSEC_small}).  However, \textsc{Qepcad} also returns an error message
\begin{verbatim}
Error! Delineating polynomial should be added over cell(2,2)!
\end{verbatim}
indicating that the output is not guaranteed to be correct.  In fact the output is correct since the the error message was triggered by the nullification of $g$ when $x=y=0$ which does not invalidate the theory.  Thus it is the error message which is incorrect: no delineating polynomial was required.
\end{example}

\subsection{Smaller lifting sets}
\label{SUBSEC_small}

Traditionally in CAD, the projection phase identifies a set of projection polynomials, which are then used in the lifting phase at sample points to create the stacks.  However when constructing CADs with respect to equational constraints we can be more efficient by discarding some of the projection polynomials before lifting.  
The non-equational constraints (in the main variable) are part of the set of projection polynomials, required in order to produce subsequent projection polynomials through their resultant with the equational constraint.  However, these polynomials are not then usually  required for the lifting.  

In Algorithm \ref{alg_ECCAD} the projection polynomials are formed from the input polynomials (in the main variable) and the set of polynomials $\mathfrak{P}$ constructed in step \ref{step_EC_mfP} which are not in the main variable.  The  lower dimensional CAD $D$ constructed in step \ref{step_EC_CF} is guaranteed to be sign-invariant (in fact order invariant) for $\mathfrak{P}$.  In particular, $\mathfrak{P}$ contains the resultants of the equational constraint with the other constraints and thus $D$ is already decomposing the domain into cells such that the presence of an intersection of $f$ and $g$ is invariant in each cell.  Hence for the final lift we need only ensure that $f$ is sign-invariant. 

As demonstrated by Examples \ref{ex_EC-small} and \ref{ex_EC-small2}, using smaller lifting sets can reduce the number of cells in a CAD.

\begin{example}
\label{ex_EC-small}
Consider again the circle and hyperbola in Figure \ref{fig_exEC} and the formula $f=0 \wedge g<0$ from Example \ref{ex_EC1}.  Using Algorithm \ref{alg_ECCAD} \texttt{ProjectionCAD} produces a CAD invariant with respect to the equational constraint using 53 cells.  This may be compared to \textsc{Qepcad} which after declaring the equational constraint produces a CAD with 69 cells.  

Both implementations give the same induced CAD of the real line but \textsc{Qepcad} uses more cells for the CAD of $\R^2$.  In particular, \texttt{ProjectionCAD} has a cell where $x<-2$ and $y$ is free while \textsc{Qepcad} uses three cells, splitting where $g$ changes sign.  The splitting is necessary for a sign-invariant CAD but not a CAD with respect to an equational constraint since $f$ is non-zero for all $x<-2$.  \textsc{Qepcad} splits the cell unnecessarily since $g$ is in the set of projection polynomials and thus used in the final lift.  
\end{example}

\begin{example}
\label{ex_EC-small2}
Consider again the polynomials and formula from Example \ref{ex_TTICAD1}.  Here we explained how the problem could be tackled using an implicit equational constraint.  Using Algorithm \ref{alg_ECCAD} in \texttt{ProjectionCAD} gives a CAD with 145 cells while declaring the equational constraint in \textsc{Qepcad} results in a CAD with 249 cells.
\end{example}

\subsection{Optimising the lifting to reduce cell counts and avoid unnecessary failure}
\label{SUBSEC_risk}

In Subsection \ref{SUBSEC_small} above we described how smaller lifting sets are used to reduce cell counts.  Actually, Algorithm \ref{alg_LSet} does sometimes use the original larger lifting sets, (in steps \ref{step_EC_addB1} and \ref{step_EC_addB2}).  These exceptions occur when an equational constraint is nullified.  If this occurs over a zero dimensional cell (stet \ref{step_EC_addB1}) then we know that the sample point is certainly representative of the cell and hence can proceed to make a stack of cells such that all the polynomials are sign-invariant trivially.   

The other exception is implemented in steps \ref{step_EC_excl1} to \ref{step_EC_addB1} of Algorithm \ref{alg_LSet}. 
In step \ref{step_EC_excl1} we form ExclP$_E(A):=P(A \setminus E) \setminus P_E(A)$ which is the set of projection polynomials that would have been calculated if using the original McCallum projection operator, but which are ignored by the reduced operator.  That is,
\[
P(A) = P_E(A) \cup \rm{ExclP}_E(A).
\]  
If all the polynomials in ExclP$_E(A)$ are constant then we can use the theory of the original projection operator in \cite{McCallum98} to conclude that the output of the algorithm will be a valid CAD.  However, this requires the lifting set to contain all the projection polynomials.

Algorithm \ref{alg_ECCAD} uses Algorithm \ref{alg_LSet} to identify the best lifting set to use in each cell for the final lift.  Hence when the lifting set is extended in this way it only effects the necessary cells thus minimising the cell count while maximising the success of the algorithm.  Example \ref{ex_EC-risk} demonstrates this.

\begin{example}
\label{ex_EC-risk}
Assume the variable ordering $w>z>y>x$.  Consider the polynomials
\[
f = z+yw, \quad
g = yx+1, \quad
h = w(z+1)+1, 
\]
and the formula $f=0 \wedge g<0 \wedge h<0$.  Using the \texttt{ProjectionCAD} package we can build a full sign-invariant CAD for $\{f,g,h\}$ with 927 cells and a CAD invariant with respect to the equational constraint with 467 cells.  
The induced CAD of $\R^3$ has 169 cells and on five of these cells the polynomial $f$ is nullified.  On these five cells both $y$ and $z$ are zero, with $x$ being either $0,4$ or on the three intervals splitting at these points create.

In this example ExclP$_E(A) = \{z+1\}$ arising from the coefficient of $h$.  We see that this is a constant value of 1 on all five of the cells above.  Thus the algorithm is allowed to proceed without error, lifting with respect to all the projection polynomials on these cells.

We note that the lifting set varies from cell to cell in $D$.  For example, the stack over the cell $c_1 \in D$ where $x=y=z=0$ uses three cells, splitting when $w=-1$.  This is required for a CAD invariant with respect to $f$ since $f=0$ on $c$ but $h$ changes sign when $w=-1$.  Compare this with, for example, the cell $c_2 \in D$ where $x=y=0$ and $z<-1$.  The stack over $c_2$ has only one cell, with $w$ free.  The polynomial $h$ will change sign over this cell, but this is not relevant since $f$ will never be zero.  This is achieved by including $h$ in the lifting set only for the five cells of $D$ where $f$ was nullified.
We note that for this example \textsc{Qepcad} can make use of partial CAD trial evaluation techniques to build a CAD invariant with respect to $f$ using only 203 cells.
\end{example}

\section{Extensions to truth-table invariant CAD} 
\label{SEC_TTICAD}

Algorithm \ref{alg_TTICAD} describing the implementation of TTICAD in the \texttt{ProjectionCAD} package is an extended version of the algorithm given in \cite{BDEMW13}.  

It includes all three of the approaches for improved lifting with equational constraints discussed in Section \ref{SEC_EC}.  The first (checking the finer well-orientedness condition) and third (adapting the lifting set as required) were already used in \cite{BDEMW13} and follow automatically from the main theorem allowing the use of the reduced projection of $\mathcal{A}$ with respect to $\mathcal{E}$, (Theorem 3 in \cite{BDEMW13}).  The second (analysing the excluded polynomials to reduce the risk of failure from not being well-oriented) was also discussed in \cite{BDEMW13}, although it was not included in the algorithm there for simplicity.  Lemma 7 in \cite{BDEMW13} validates its use in the TTICAD case.   

However, the most significant extension is the relaxation of the input to allow formulae which do not have any equational constraint.  Although the extension is reasonably straightforward, it actually dramatically increases both the applicability and benefit of TTICAD .

\subsection{Extending the applicability of TTICAD}

Algorithm \ref{alg_TTICAD} allows the user to input a sequence of formulae, each of which may or may not have an equational constraint.  This is in contrast to the theory in \cite{BDEMW13} where it was assumed each formula had one.  In Algorithm \ref{alg_TTICAD} formulae without equational constraints are dealt with by treating all their polynomials with the importance reserved for equational constraints, (step \ref{step_T1} to \ref{step_T3}).  

The main theorem of \cite{BDEMW13} validating the use of the projection operator (Theorem 3) requires no more adjustment that the redefinition of $E_i$ (and thus $E$ and $\mathcal{E}$) introduced by these steps.  Extra polynomials have been added to the projection set sufficient to allow the conclusion that: (a) the CAD is sign-invariant with respect to all the equational constraints and all the other constraints belonging to a formulae with no equational constraint, and (b) in cells where an equational constraint vanishes the other constraints its formulae are sign invariant.  

However, as with Theorem 3 in \cite{BDEMW13} this theorem holds when the equational constraints are not nullified and thus the well-orientedness condition must be extended to include the redefined $\mathcal{E}$ from the algorithm.  This is indeed the case in the implementation since when a clause without an equational constraint has $E_i$ redefined accordingly in step \ref{step_T2} and this set is passed to Algorithm \ref{alg_LSet} where it is used in the check at step \ref{step_EC_WO}.

\subsection{Increasing the benefit of TTICAD}

In this final section we demonstrate how the extension dramatically increases the benefit of the TTICAD theorem.  
First we note that in \cite{BDEMW13} experimental results demonstrated the great benefit of TTICAD over sign-invariant CADs through much smaller cell counts.  The benefits increase in proportion to the number of formulae, as in Example \ref{ex_TTICAD2}.

\begin{example}
\label{ex_TTICAD2}
We consider a family of examples based on Example \ref{ex_TTICAD1}.  Define the $f_1,g_1,f_2,g_3$ as in that example and consider also
\[
f_3 = (x+4)^2+(y+1)^2-1, \qquad
g_3 = (x+4)(y+1)-1/4.
\]
Then consider the formulae
\begin{align*}
\Phi_1&= \left(f_1 = 0 \land g_1 < 0 \right), \qquad 
\Phi_2= \left(f_1 = 0 \land g_1 < 0 \right)\lor \left( f_2 = 0 \land g_2 < 0   \right), \\
\Phi_3&= \left(f_1 = 0 \land g_1 < 0 \right)\lor \left( f_2 = 0 \land g_2 < 0   \right)\lor \left( f_3 = 0 \land g_3 < 0   \right).
\end{align*}
For each formula we used \texttt{ProjectionCAD} to construct a full sign-invariant CAD for the polynomials, a CAD invariant with respect to the implicit equational constraint and a TTICAD, using Algorithm 3 in \cite{England13} and Algorithms \ref{alg_ECCAD} and \ref{alg_TTICAD} from this report respectively.  The sequence of formulae for TTICAD is the obvious one (breaking the formula at the disjunctions).  We also used \textsc{Qepcad} to analyse the formula on both its default settings and by declaring the implicit equational constraint manually.  Table \ref{tab_TTICAD} shows the cell counts from these experiments.

\begin{table}[hb]
\centering
\begin{tabular}{|c|ccc|cc|}
\hline
\textbf{Formula}    & \multicolumn{3}{|c|}{\texttt{ProjectionCAD}} & \multicolumn{2}{|c|}{\textsc{Qepcad}} \\
& \texttt{CADFull} & \texttt{ECCAD} & \texttt{TTICAD} 
& \texttt{Default} & \texttt{Declare}  \\
\hline
$\Phi_1$ & 83  & 69  & 53  & 69  & 83  \\
$\Phi_2$ & 317 & 145 & 105 & 317 & 249 \\ 
$\Phi_3$ & 695 & 237 & 157 & 695 & 509    \\  
\hline
\end{tabular}
\caption{Table detailing the number of cells in CADs constructed using various algorithms to analyse the formulae in Example \ref{ex_TTICAD2}.}
\label{tab_TTICAD}
\end{table}

As expected, the TTICAD is the superior of these algorithms in terms of cell counts, with its advantage increasing with the number of disjunctions (separate formulae in the sequence inputted to the algorithm.  \textsc{Qepcad} can also use the theory of equational constraints to avoid building a full sign-invariant CAD (although it requires manual input), however, the smaller lifting sets used in \texttt{ProjectionCAD} (Subsection \ref{SUBSEC_small}) mean that \textsc{Qepcad} has higher cell counts.
\end{example}

The benefit of the extended TTICAD over a sign-invariant CAD will now increase in proportion with the number of the formulae \textit{that have an equational constraint}.  Hence, for such relaxed input, the benefit over sign-invariant CAD will be slightly less.  However, for such input, the theory of equational constraints alone is not applicable (as there is no overall implicit equational constraint).  Thus in these cases TTICAD is the only available theory to consider the structure given by the equations, and therefore of much greater importance.  Example \ref{ex_TTICAD3} demonstrates this.

\begin{example}
\label{ex_TTICAD3}
Consider again the polynomials from Example \ref{ex_TTICAD2} and the formulae $\Phi_1, \Phi_2, \Phi_3$ but this time with $f_1<0$ instead of $f_1=0$.  For these formulae TTICADs can be produced with 83, 183 and 283 cells respectively.  Although a little larger than the cell counts in Table \ref{tab_TTICAD}, these still represent great savings over the full sign-invariant CADs.  Further, for these modified formulae the \texttt{ECCAD} algorithm and declaring an equational constraint in \textsc{Qepcad} is not applicable, making the difference between the possible cell counts for each problem much more dramatic.
\end{example}

\section{Summary}
\label{SEC_Summary}

We have described the new functionality present in the second release of \texttt{ProjectionCAD}.  This includes an algorithm to produce a CAD invariant with respect to an equational constraint.  Similar algorithms are present in other implementations such as \textsc{Qepcad}, but these usually just include improvements to the projection phase, resulting in a smaller projection set and hence a CAD with less cells.  In Section \ref{SEC_EC} we described how the theory of equational constraints also allows for improvements to the lifting phase which can further reduce the cell counts.  

The second release also includes an algorithm to produce truth-table invariant CADs, the only such algorithm to be implemented.  This is based on the theory in \cite{BDEMW13} but in Section 4 we describe how that theory can be extended to increase the applicability of the algorithm to situations where its benefit is greatest.  

Throughout this report we have focussed on cell counts as the measure of the success of the implementation.  Of course, the actual time taken to compute the CADs is also important and it is the case that \textsc{Qepcad} and the other state of the art algorithms (\texttt{RegularChains} in Maple \cite{CMXY09} and Mathematica \cite{Strzebonski10}) are usually quicker to produce sign-invariant CADs.  However, as reported in \cite{BDEMW13} the implementation of TTICAD offers such large reductions in cell counts that it allows \texttt{ProjectionCAD} to compete on timings as well as cell counts.  The extension to TTICAD described in this report will further increase the relative performance of TTICAD since it can now be applied to examples where the theory of equational constraints alone is no use.  

Finally, we note that the second release of \texttt{ProjectionCAD} includes implementations of the work in \cite{BDEW13} on heuristics for choosing how to formulate problems for the algorithms.  The problem formulation can have a dramatic effect on the computation and future work on \texttt{ProjectionCAD} will include automating the use of these heuristics within the main algorithms. 
  
\begin{footnotesize}
\bibliography{CAD}{}
\bibliographystyle{plain}
\end{footnotesize}

\end{document}